\documentclass{article}
\usepackage{amsmath}
\usepackage{amssymb}
\newcommand{\df}[2]{\mbox{$\frac{#1}{#2}$}}
\newtheorem{Lem}{Lemma}
\newtheorem{Thm}{Theorem}
\newtheorem{Eg}{Example}
\newtheorem{Conj}{Conjecture}
\newcommand{\mcl}{\operatorname{mcl}}
\newcommand{\mgl}{\operatorname{mgl}}
\newcommand{\li}{\operatorname{Li}}
\newcommand{\eop}{\vrule height .9ex width .9ex depth -.1ex}

\begin{document}

\title{\bf Central Binomial Sums, Multiple Clausen Values and Zeta
Values}

\author{Jonathan Michael Borwein, FRSC\\
Shrum Chair of Science\\
Department of Mathematics and Statistics\\
Simon Fraser University, Burnaby, BC V5A 1S6 Canada\\
email: jborwein@cecm.sfu.ca
\and
David J. Broadhurst \\
Physics Department \\
Open University, Milton Keynes, MK7 6AA, UK\\
e-mail: D.Broadhurst@open.ac.uk
\and
Joel Kamnitzer\\
Department of Applied Mathematics\\
University of Waterloo, Waterloo, ON N2L 3G6 Canada\\
email: jkamnitz@uwaterloo.ca
}

\date{3 April 2000}

\maketitle

\begin{abstract}\noindent
We find and prove relationships between Riemann zeta values and central
binomial sums. We also investigate alternating binomial sums (also called
Ap\'ery sums). The study of non-alternating sums leads to an investigation of
different types of sums which we call multiple Clausen values. The study of
alternating sums leads to a tower of experimental results involving
polylogarithms in the golden ratio. In the non-alternating case, there is a
strong connection to polylogarithms of the sixth root of unity, encountered
in the 3-loop Feynman diagrams of {\tt hep-th/9803091} and subsequently in
{\tt hep-ph/9910223}, {\tt hep-ph/9910224}, {\tt cond-mat/9911452} and {\tt
hep-th/0004010}.
\end{abstract}

\vspace{\baselineskip}
\vfill
\noindent{\bf AMS (1991) subject classification}: Primary 40B05, 33E20,
Secondary 11M99, 11Y99.

\vspace{2true cm}

\section{Introduction}

We shall begin by studying the {\em central binomial sum} $ S(k) $, given as:
\begin{equation}
\label{eq:bin}
S(k) := \sum_{n=1}^\infty \frac{1}{n^k \,{2n \choose n}}
\end{equation}
for integer $k$.
A classical evaluation is $S(4)=\frac{17}{36} \zeta(4)$.  Using a
mixture of integer relation and other computational techniques, we
uncover remarkable links to values multi-dimensional polylogarithms of
sixth roots of unity which we call multiple Clausen values.  We are thence
able to prove some surprising identities -- and\ empirically determine
many more. Our experimental {\em integer relation} tools are
described in some detail in~\cite{BL}.

 We shall finish by discussing the corresponding
alternating sum:
\begin{equation}
\label{eq:alt}
A(k):=\sum_{n=1}^\infty \frac{(-1)^{n+1}}
{n^k\,{{2n}\choose n}}.
\end{equation}
These are related to {\em polylogarithmic
ladders} in the {golden ratio} $\frac {\sqrt 5 -1}2$.  A classical
evaluation is $A(3)=\frac{2}{5} \zeta(3)$, with its connections to
Ap\'ery's proof of the irrationality of $\zeta(3)$, (see e.g.,~\cite{PA}).

\section{Definitions and Preliminaries}

We start with some definitions which for the most part follow Lewin~\cite{L},
and~\cite{BBB,BBBLC,BBBL}.
A useful {\em multi-dimensional polylogarithm} is defined by
$$ \li_{a_1, \dots , a_k} (z) :=
\sum_{n_1 > \dots > n_k > 0} \frac{z^{n_1}}{n_1^{a_1} \ldots n_k^{a_k}}, $$
with the parameters required to be positive integers.
  This is a generalization of the familiar polylogarithm
   $ \li_n(z) := \sum_{k=1}^\infty z^k/k^n $.
Note that $ \li_n(1) = \zeta(n) $.

We will be most concerned with the value of the
multi-dimensional polylogarithm at the sixth root of unity,
 $ \omega := e^{i\pi/3} $.  We refer to such a value as a
  {\em multiple Clausen value}  (MCV)  and write
  $$ \mu(a_1, \dots , a_k)
  := \li_{a_1,  \dots , a_k} (\omega).$$
    This MCV is analogous to the
multiple zeta value (MZV), at $z=1$, which has been studied
 in works such as~\cite{BBB,BBBLC,BBBL}.  We might also have viewed these as
generalizations of the {\em Lerch zeta} function.
  It transpires to be
  advantageous to separate the real and
 imaginary parts of an MCV in a manner
 that is based on the sum of the arguments.
 We refer to these parts as {\em multiple Glaishers}
  (mgl) and {\em multiple Clausens} (mcl).  They are defined by:
\begin{align*}
\mgl(a_1, \dots , a_k) := &Re(i^{a_1 + \cdots + a_k}\mu(a_1, \dots, a_k)) \\
\mcl(a_1, \dots , a_k) := &Im(i^{a_1 + \cdots + a_k}\mu(a_1, \dots,
a_k)),
\end{align*}
and may be written explicitly as multiple $\sin$ or $\cos$ sums
depending on
the parity. For example, when $a+b$ is odd,
$$\mgl(a,b)=\pm \sum_{n>m>0}\frac{\sin(n\frac{\pi}3)}{n^am^b},$$
as is the case in Theorem 1 below.

As elsewhere, the {\em weight} of a sum is  $\sum_{i=i}^k a_i$
while the {\em depth} $k$ is the number of parameters.
This separation corresponds, in the case $k=1$,
to Lewin's (\cite{L}) separation of the polylogarithm
 at complex exponential arguments into Clausen and
  Glaisher functions.

We record the following differential properties of our
multi-dimensional polylogarithm:
\begin{align}
\frac{d\li_{a_1 + 2, a_2, \dots , a_n}(z)}{dz}
&= \frac{\li_{a_1 + 1, \dots , a_n}(z)}{z},\label{eq:lid1}\\
\frac{d\li_{1, a_2, \dots , a_n}(z)}{dz}
&= \frac{\li_{a_2, \dots , a_n}(z)}{1-z}. \label{eq:lid2}
\end{align}
Repeated application of (\ref{eq:lid2}) yields:
\begin{equation}
\label{eq:lid3}
\li_{ \{ 1 \}^n}(z) = \frac{(-\log(1-z))^n}{n!},
\end{equation}
where $ \{ 1 \}^n $ denotes the string $ 1, \dots , 1 $ with $ n $ ones.

We will make  some use of the {\em Bernoulli polynomials} later in this paper.
 Recall that $ B_n(x) $ is defined by
 $$ \frac{te^{xt}}{e^t -1} = \sum_{n=0}^\infty B_n(x) \frac{t^n}{n!},$$
 and  that  $ B_n(0) $ is called the {\em nth Bernoulli number} and is
 written $B_n$.

For convenience we choose the following notation for {\em log-sine integrals}.
 We define
\begin{equation}
\label{eq:j}
 j(a,b) := \int_0^{\pi/3} (\log(2\sin \frac{\theta}{2} ))^a
 \theta^b d\theta
 \end{equation}
\begin{equation}
\label{eq:r}
  r(a,b) := \frac{i^{b+1}}{a!b!} \int_0^{\pi/3}
 \left( \log \left( 2\sin \frac{\theta}{
2} \right) + \frac{i}{2}\left( \theta - \pi \right) \right)^a \theta^b
d\theta .\end{equation}

Finally, note the following standard result involving the {\em Gamma
function}
which will prove very useful:
$$
a^{-n}\Gamma(n) = \int_0^1 y^{a-1} (-\log y)^{n-1} dy.
$$

\section{Non-alternating Central Binomial Sums}

Our first step is to write $ S(k) $ in integral form.
\begin{Lem} For all  positive integers
\begin{equation}
 S(k) = \frac{(-2)^{k-2}}{(k-2)!} j(k-2, 1).
\end{equation}
\end{Lem}

{\raggedright \bf Proof.}
Employing the Gamma function and various standard tricks, we have:
\begin{align*}
S(k) &= \sum_{n=1}^\infty \frac{1}{{2n \choose n} n^k}  \\
	    &= \sum_{n=1}^\infty \frac{1}{{2n \choose n} \Gamma(n)}
	     \int_0^1 (-\log x)^{k-1} x^{n-1} dx  \\
	    &= \sum_{n=1}^\infty \frac{1}{{2n \choose n} \Gamma(n)}
	     \int_0^1 (-2\log y)^{k-1} y^{2n -2} (2y) dy
	     \quad \text{(by $ y^2 = x $)} \\
	   &=- \sum_{n=1}^\infty \frac{(-2)^{k-1}}{ {2n \choose n}
	   \Gamma(n)} \left(
	    \int_0^1 \frac{y^{2n}}{2n} 2(k-1)\frac{(\log y)^{k-2}}{y} dy
	    \right)\\
	     &\quad \text{(after integrating by parts)}  \\
	  &= \frac{(-2)^{k-1}}{(k-2)!} \int_0^1
\frac{(\log y)^{k-2}}{y} \sum_{n=1}^\infty \frac{y^{2n}}{n {2n \choose n}}dy\\
	 &= \frac{(-2)^{k-1}}{(k-2)!} \int_0^1 \frac{(\log y)^{k-2}}{y}
	 \frac{y        \arcsin \frac{y}{2}}{\sqrt{1-(\frac{y}{2})^2}} dy
	 \quad  \text{(see~\cite{PA} p. 384)} \\
	&= \frac{(-2)^{k-2}}{(k - 2)!} \int_0^{\pi/3} (\log(2 \sin
	\frac{\theta}{2}))^{k-2} \theta d\theta
	\quad \text{(by $ y = 2\sin \frac{\theta}{2} $)}.
\end{align*}
$\eop$

\vspace{\baselineskip}

Hence to evaluate the sums $ S(k) $, it is enough to determine log-sine
 integrals of the {\it special} form $ j(k,1) $.
 The key is the following identity:

\begin{Lem} For all $k > 2$,
\begin{equation}
 \sum_{r=0}^{k-2} \frac{(-i\pi/3)^r}{r!} \mu(k-r, \{ 1 \}^n)
 = \zeta(k, \{ 1 \}^n) - (-1)^{k+n}r(n+1, k-2). \label{eq:zp}
\end{equation}
\end{Lem}

{\raggedleft \bf Proof.}
First note the formal identity:
\begin{equation}
\log (1 - e^{i\theta}) = \log \left( 2 \sin \frac{\theta}{2} \right)
+ \frac{i}{2}(\theta - \pi).\label{eq:h1}
\end{equation}
We have
$$
\mu(k, \{ 1 \}^n) = \zeta(k, \{ 1 \}^n)
 + \int_1^\omega \frac{\li_{k-1, \{ 1 \}^n}(z)}{z} dz \quad \text{(by
 (\ref{eq:lid1}))}.
$$
We now integrate by parts repeatedly to obtain:
\begin{equation*}
\begin{split}
\mu(k, \{ 1 \}^n) &= \zeta(k, \{ 1 \}^n) -
\sum_{r=1}^{k-2} \frac{(-1)^r \mu(n-r, \{ 1 \}_k) \log^r(\omega)}{r!} \\
 &+ \frac{(-1)^{k-2}}{(k-2)!} \int_1^\omega \frac{\li_{\{ 1 \}_{n+1}} (z)
 \log^{k-2}(z)}{z}
 dz.
\end{split}
\end{equation*}

Observe that $\log(\omega) = \frac{i\pi}{3} $.  Let $ z = e^{i\theta} $.
  Then we have (using (\ref{eq:lid3})):
\begin{equation*}
\begin{split}
\mu(k, \{ 1 \}^n) &= \zeta(k, \{ 1 \}^n) - \sum_{r=1}^{k-2}
\frac{(-i\pi/3)^r \mu(n-r, \{ 1 \}_k)}{r!} \\
&+ \frac{i(-1)^{k}}{(k-2)! (n+1)!}
\int_0^{\frac{\pi}{3}}\ (-\log(1-e^{i\theta}))^{n+1} (i\theta)^{k-2}
d\theta,
\end{split}
\end{equation*}
which gives us the desired result after applying (\ref{eq:h1}).
\eop

\vspace{\baselineskip}

Now, note that $ r(a,b) $, defined in (\ref{eq:r}), can be expanded out
 binomially and
written
as linear
 rational combination of $ j(c,d) $, defined in (\ref{eq:j}), for various
 $ c, d $, including
 a non-zero multiple of $ j(a, b) $.
 So we may repeatedly use the above identity to solve for each
 $ j(a,b) $ in terms of multiple Clausen, Glaisher and Zeta values -- all
 of the same form
 $ \mcl(n, \{ 1 \}_k)$, $ \mgl(n, \{ 1 \}_k)$ and $\zeta(n, \{ 1 \}_k) $.

In particular for all $ k $, $ j(k-2, 1) $ and hence $ S(k) $ can be
 written as a linear rational combination of multiple zeta, Clausen and Glaisher
  values of this form.
   This method (which we have automated in both {\em Reduce} and {\em Maple})
    recovers all previously known results in a uniform fashion.
    It does not in general give especially nice looking identities,
    but we are able to apply some other results about multiple Clausen values
    derived in the next section to clean things up for small $ k $.
     After doing this, we obtain the following results:

\begin{Thm} The following evaluations of central binomial sums hold:
\begin{align*}
S(2) &= \frac{\zeta(2)}{3}  \\
S(3) &= - \frac{2\pi}{3} \mcl(2) - \frac{4}{3} \zeta(3) \\
S(4) &= \frac{17}{36} \zeta(4) \\
S(5) &= 2 \pi \mcl(4) - \frac{19}{3}\zeta(5) + \frac{2}{3}\zeta(3)\zeta(2) \\
S(6) &= - \frac{4\pi}{3} \mgl(4,1) + \frac{3341}{1296}\zeta(6)
-       \frac{4}{3}\zeta(3)^2 \\
S(7) &= -6 \pi \mcl(6) - \frac{493}{24}\zeta(7) + 2\zeta(5)\zeta(2)
+ \frac{17}{18}\zeta(4)\zeta(3) \\
S(8) &= -4 \pi \mgl(6,1) + \frac{3462601}{233280}\zeta(8) -
\frac{14}{15}\zeta(5,3) -
\frac{38}{3}\zeta(5)\zeta(3) + \frac{2}{3}\zeta(2)\zeta(3)^2.
\end{align*}
\end{Thm}

Note that the results for $ S(2) $ and $ S(4) $ are classical evaluations.
The others are  new and it is hoped that they will shed light on odd
$\zeta$-values, which remain a  source of many unanswered questions. We
observe that genuine MCVs. with depth $k>1$,   first occur for $n=6$ and
8.  Moreover, David Bailey and David Broadhurst have explicitly
obtained $S(n)$ for $n\leq20$ through   a very high level application of
integer relation algorithms. The result for $S(20)$ is presented in
\cite{DHJB1}.

\section{Multiple Clausen Values}

Central binomial sums naturally led us into a study of multiple Clausen values.
 This study proved to be quite fruitful and we were led to many
  striking results.

To start with, we quote the following results about depth-one Clausen values
as described  in~\cite{L}:
\begin{gather}
\mgl(n) = \frac{(-1)^{n+1}2^{n-1}\pi^n B_n(\frac{1}{6})}{n!}, \label{eq:l1} \\
\mcl(2n+1) = \frac{(-1)^n}{2}(1-2^{-2n})(1-3^{-2n})\zeta(2n+1). \label{eq:l2}
\end{gather}

\subsection{MCV Duality}

In MZV analysis, one of the central results is the now well-known MZV
 {\em duality theorem} recapitulated in~\cite{BBB}:
\begin{equation}
\label{eq:mzdu}
\zeta(a_1 + 2,\{ 1 \}_{b_1}, \dots, a_k + 2, \{ 1 \}_{b_k})
= \zeta(b_k + 2,\{1\}_{a_k}, \dots b_1+2,\{1\}_{a_1}),
\end{equation}
 For all positive integers $a_1, a_2, \dots, a_n$.

For MCV's, we have found two such duality results,
with the first result applying if the first argument of the MCV is
one
(such sums converge as in the classical Fourier setting)
while the second holds if the first argument of the MCV is
two or more.
The pattern  in Theorem \ref{mcvd1}
is somewhat complicated.  A prior  example may make things clearer.

\begin{Eg}
\begin{equation*}
\begin{split}
&\mu(1,3,1,2) - \mu(1)\mu(3,1,2) + \mu(2)\mu(2,1,2) - \mu(3)\mu(1,1,2)
+ \mu(1,3)\mu(1,2) \\
&\quad -\mu(1,1,3)\mu(2) + \mu(2,1,3)\mu(1) - \mu(1,2,1,3) = 0.
\end{split}
\end{equation*}
\end{Eg}

Note that each term differs from the previous by subtracting `1' from
the first argument of the right MCV and adding `1' to the first
argument of the left MCV.  In the case where there is a `1' as the
first argument of the right MCV, this `1' is dropped and concatenated
onto the left MCV.

\begin{Thm} \label{mcvd1} For all positive integers $a_1, a_2, \dots, a_n$
\begin{equation}
\begin{split}
\label{eq:mcdu2}
&\quad \mu(1,a_1, \dots, a_n) - \mu(1)\mu(a_1, \dots, a_n)
+ \mu(2)\mu(a_1 - 1, \dots, a_n) + \dots \\
&\pm \mu(a_1)\mu(1, a_2, \dots , a_n) \mp
\mu(1,  a_1)\mu(a_2, \dots , a_n) + \dots \pm \mu(1, a_n, ..., a_1) =
0.
\end{split}
\end{equation}
\end{Thm}

{\raggedleft \bf Proof.}
We will prove this by repeated integration by parts.
We use the differential properties (\ref{eq:lid1})
and (\ref{eq:lid2}) to move weight from one multidimensional
polylogarithm to another:
\begin{align*}
\mu(1, a_1, \dots, a_n) &= \int_0^\omega \frac{\li_{a_1, \dots, a_n}(z)}
{1-z}dz ,\\
&= \left[ \li_1(z) \li_{a_1, \dots, a_n}(z) \right]_0^\omega -
\int_0^\omega \frac{\li_1(z) \li_{a_1-1, \dots, a_n}(z)}{z}dz \\
&\dots \\
&=\dots \pm \mu(1, a_n, \dots, a_1).
\end{align*}
\eop

\vspace{\baselineskip}

As in other duality results, it is interesting to examine what happens in
the self-dual case.  Suppose that $ (a_1, \dots, a_n) = (a_n, \dots,
a_1) $, then if $ a_1 + \dots + a_n $ is even, (\ref{eq:mcdu2}) reduces
to $ 0 = 0 $.  If the sum is odd, then (\ref{eq:mcdu2}) shows that
$\mu(1, a_1, \dots, a_n) $ reduces to a sum and product of lower
weight MCVs.

The pattern in Theorem \ref{mcvd2} below is somewhat more complicated.
Hence an
example will be even more instructive.  (The bar denotes complex
conjugation.)

\begin{Eg}
\begin{equation*}
\begin{split}
&\mu(4,3,1) + \overline{\mu(1)}\mu(3,3,1) + \overline{\mu(1,1)}\mu(2,3,1)
+ \overline{\mu(1,1,1)}\mu(1,3,1) \\
&\quad + \overline{\mu(2,1,1)}\mu(3,1) + \overline{\mu(1,2,1,1)}\mu(2,1)
 + \overline{\mu(1,1,2,1,1)}\mu(1,1) \\
&\quad + \overline{\mu(2,1,2,1,1)}\mu(1) + \overline{\mu(3,1,2,1,1)} =
\zeta(3,1,2,1,1).
\end{split}
\end{equation*}
\end{Eg}

Note that each term differs from the previous by subtracting  `1'
 from the first argument of the right MCV and concatenating `1' onto
 the left MCV.  In the case where there is a 1 as the first argument of
 the right MCV, this `1' is dropped and  `1' is added to the first
 argument of the left MCV .

Theorem \ref{mcvd2} is  specialization of the {\em H\"older convolution} ((44)
from~\cite{BBBL}) with $ p = \omega $ and $ q = 1 - \omega =
\overline{\omega}$.   That paper gives a  more formal description of
the pattern of summation that we have outlined above.

\begin{Thm} \label{mcvd2} For all positive integers $a_1, a_2, \cdots, a_n$
\begin{equation}
\begin{split}
\label{eq:mcdu1}
&\mu(a_1 + 2,\{ 1 \}_{b_1}, \dots, a_k + 2, \{ 1 \}_{b_k}) +
\overline{\mu(1)}\mu(a_1 +      1,\{ 1 \}_{b_1}, \dots, a_k + 2, \{ 1
\}_{b_k})       + \dots + \\
&\overline{\mu(\{ 1 \}_{a_1 +   1})}\mu(1,\{ 1 \}_{b_1}, \dots, a_k + 2,
\{ 1 \}_{b_k})
+ \overline{\mu(2, \{ 1 \}_{a_1})}\mu(\{ 1 \}_{b_1}, \dots, a_k + 2,
\{      1 \}_{b_k}) \\
&+ \dots + \overline{\mu(b_1 + 2, \{ 1 \}_{a_1})}\mu(a_2 + 2, \dots, a_k +
2, \{ 1 \}_{b_k}) + \dots + \\
&\overline{\mu(b_k + 2,\{1\}_{a_k}, \dots b_1+2,\{1\}_{a_1})}
= \zeta(b_k + 2,\{1\}_{a_k}, ... b_1+2,\{1\}_{a_1})
\end{split}
\end{equation}
\end{Thm}

{\raggedright \bf Proof.}
As stated above, this follows by  H\"older convolution,
 since $ \li_{\vec{a}}(\overline{\omega}) = \overline{\mu(\vec{a})}$.
 It can also be proved using a similar integration by parts as
  above, after noting that the following differentiation result holds:
\begin{align*}
\frac{d\li_{a_1 + 2, a_2, \dots, a_n}(1-z)}{dz} &
= -\frac{\li_{a_1 + 1,  a_2, \dots, a_n }(1 - z)}{1 - z} \\
\frac{d\li_{1, a_2, \dots, a_n}(1-z)}{dz} &= -\frac{\li_{a_2, \dots, a_n }(1 -
z)}{z}.
\end{align*}

This identity allows us to move weight from a multi-dimensional polylogarithm
at $ z $ to one at $ 1 - z $.
\eop

\vspace{\baselineskip}
Note that this suggested integration by parts
 technique also yields a new proof for this type of H\"older
 convolution.
  Hence, it provides another proof of MZV duality.

In this case, self-dual strings do not tell us that much.  The only thing
to note is that all the imaginary terms on left side of (\ref{eq:mcdu1})
will vanish, since $ a\overline{b} + \overline{a}b $ is real.
When $ k = 1 $, (\ref{eq:mcdu1}) simplifies to the following, on using
(\ref{eq:l1}),
\begin{equation}
\begin{split}
\label{eq:du3}
\zeta(a+2, \{ 1 \}_b) &+
\frac{(-1)^a (i\pi/3)^{a+b+2}}{(a+1)!(b+1)!} = \\
&\sum_{r=0}^a \frac{( -i\pi/3 )^r}{r!} \mu(a + 2 - r, \{ 1 \}_b)
+ \sum_{r=0}^b \frac{(i\pi/3)^r}{r!} \overline{\mu(b + 2 - r, \{ 1
\}_a)}.
\end{split}
\end{equation}
We initially derived (\ref{eq:du3}) by means of (\ref{eq:zp}) and the following
satisfying identity involving log-sine integrals, which we proved
using contour
integration:
\begin{equation}
\label{eq:jz}
(-1)^{a+b}\zeta(a+2, \{ 1 \}_{b-1}) =
 \frac{(i\pi/3)^{a+1}(-i\pi/3)^b}{(a+1)!b!} -
 \overline{r(a + 1, b - 1)} - r(b,a).
\end{equation}

\subsection{Special values of MCVs}

To illustrate the utility of this last duality result,
consider (\ref{eq:du3}) when $ a = 1 $ and $ b = 1 $.  We obtain:
$$
\zeta(3,1)-2\mgl(3,1)-\frac{2 \pi}{3} \mgl(2,1)-\frac{\pi^4}{324} =
0.
$$
Using MZV analysis~\cite{BBBL}, we know that
$ \zeta(3,1) = \frac{\pi^4}{360} $,
which is the first case in an infinite series of evaluations
in terms of powers of $\pi^2$, conjectured by Zagier and proved
in~\cite{BBBLC}.
Now from (\ref{eq:m21})    below we have $$ \mgl(2,1) = \frac{\pi^3}{324}
.$$
   This rewards us with $$ \mgl(3,1) = \frac{-23}{19440}\pi^4 .$$

Next, let us use the duality result to extract some more general evaluations.
Let
$$ F(x,y) := \sum_{a, b \ge 0} \mu(a + 2, \{ 1 \}_b) x^{a+1}
y^{b+1}.$$
According to~\cite{BBBL}, we know that this generating function is {\em
hypergeometric}:
\begin{equation}
\label{eq:gfo}
F(x,y) = 1 - {}_2F_1(-x, y; 1-x; \omega).
\end{equation}

Unfortunately, this is not a very convenient equation for
extracting coefficients or proving formulas.
To obtain a more useful representation, we take
 (\ref{eq:du3}), multiply through by $ x^{a+1} y^{b+1} $
 and sum over all $ a, b \ge 0 $.
   This gives:
\begin{equation}
\label{eq:gf1}
e^{-i\pi x/3} F(x,y) + e^{i\pi y/3} \overline{F(y,x)}
 + (e^{-i\pi x/3} - 1)(e^{i\pi y/3} -1) = G(x,y),
\end{equation}
where
$$G(x,y) := \sum_{a, b \ge 0} \zeta(a + 2, \{ 1 \}_b) x^{a+1} y^{b+1}$$
is the generating function for the corresponding  MZVs.
Now from  prior work on MZVs (see~\cite{BBB}) it is known that
$$
G(x,y) = 1 - \exp \left( \sum_{k \ge 2} \frac{x^k + y^k - (x+y)^k}{k}\zeta(k)
\right).
$$
We shall use this generating function identity to obtain more general
results about special values
 of mgl's and mcl's. First we put the last identity in a more symmetric
 form by        letting
  $ M(x,y) :=$
  $$ F(ix, -iy) = \sum_{a,b \ge 0} (-1)^{b+1} (\mgl(a + 2, \{ 1 \}_b)
  + i \mcl(a + 2, \{ 1 \}_b)) x^{a+1} y^{b+1} .$$
  Then we have:
\begin{equation}
\label{eq:gf2}
e^{\pi x /3 } M(x,y) + e^{\pi y /3 } \overline{M(y,x)}
+ (e^{\pi x /3 } - 1)(e^{\pi y /3 } -1) = G(ix,-iy).
\end{equation}

\begin{Thm} For non-negative integers $a$ and $b$
\begin{equation}
\label{eq:m21}
\mgl(\{ 1 \}_a, 2, \{ 1 \}_b) =
(-1)^{a+b+1}\frac{(\frac{\pi}{3})^{a+b+2}}{2(a+b+2)!},
\end{equation}
and depends only on $a+b$.
\end{Thm}

{\raggedright \bf Proof.}
First we show that
$$ \mgl(2, \{ 1 \}_b) =
(-1)^{b+1}\frac{(\frac{\pi}{3})^{b+2}}{2(b+2)!}.$$
We give two proofs of this result.

\medskip

\noindent {\bf (i)}
Multiply by $y/x $ in
(\ref{eq:gf2}),
and let $x$ go to zero. Set
$ \displaystyle{ B(y) := \sum_{b=0}^\infty (-1)^{b+1} \mgl(2, \{ 1 \}_b) y^{b+2}}$,
$ \displaystyle{ C(y): = \sum_{b=0}^\infty (-1)^{b+1}\mgl(b+2)y^{b+2} }$
and
$ \displaystyle{ D(y) := \sum_{m = 1}^\infty (-1)^{2m+m-1} \zeta(2, \{ 1 \}_{2m-2}) y^{2m}
}$.
Comparing real parts on each side of (\ref{eq:gf2}) yields
$$
B(y) + e^{\pi y /3}C(y) + \frac{\pi y}{3}(e^{\pi y/3 } -1) = D(y).
$$
The  value of $ \mgl(n)$ and so of $C(y)$ is given by
(\ref{eq:l1}), while MZV duality yields $ \zeta(2, \{ 1 \}_{2m-2}) = \zeta(2m)$
  with its familiar Bernoulli number evaluation.
  Combining these results and using the generating function for the Bernoulli
   polynomials, we arrive at:
\begin{align*}
B(y) = \frac{e^{\pi y /3}}{2} - \frac{\pi y}{6} - \frac{1}{2}
\Longrightarrow \mgl(2, \{ 1 \}_b) =
(-1)^{b+1}\frac{(\frac{\pi}{3})^{b+2}}{2(b+2)!}.
\end{align*}

\medskip

\noindent {\bf (ii)} Alternatively, we start with (\ref{eq:gfo}).
  Dividing by $x$, setting $ y = iy $, and letting $x$ go to zero
  yields:
$$
-i \sum_{n=1}^\infty \frac{(iy)_n \omega^n}{n n!}
=\frac{ \sum_{b = 0}^\infty \mu(a+2, \{ 1 \}_b) (iy)^{b+2}}{y}.
$$
Now we know that
$$
\sum_{n=1}^\infty \frac{(iy)_n z^{n-1}}{n!} = \frac{(1-z)^{-iy}
-1}{z}.
$$
Integrate both sides of this expression from
 $ 0 $ to $ \omega $
 along $ z = 1 + e^{i(\pi - \theta)}, \; \frac{2\pi}{3} \le \theta \le \pi
 $, to obtain,
$$
\sum_{n=1}^\infty \frac{(iy)_n \omega^n}{n \; n!} =
-i \left( \int_0^{\frac{\pi}{3}} \frac{(e^{-\theta y}- 1)(1 - \cos \theta +
 i \sin \theta)}{2 - 2 \cos \theta} d\theta + \frac{e^{-\pi y/3}}{y}
  - \frac{1}{y} + \frac{\pi}{3} \right).
$$
This again gives us:
$$
Im \left( \sum_{n=1}^\infty \frac{(iy)_n \omega^n}{n \; n!} \right)
= - \frac{e^{-\pi y /3} + \frac{\pi y}{3} - 1}{2y}
$$
as required.

\medskip

Armed with this special case, we now prove the full result
for $ \mgl(\{ 1 \}_a, 2, \{ 1 \}_b) $.
 We
start with a similar integration by parts as for the proof of (\ref{eq:zp}).
We have:
\begin{equation*}
\begin{split}
\mu(\{ 1 \}_b, 2, \{ 1 \}_a) &= \int_0^\omega \frac{\li_{ \{ 1 \}_{b-1}, 2,
\{ 1 \}_a} (z)}{1-z} dz \\
&= \frac{i \pi}{3} \mu(\{ 1 \}_{b-1}, 2, \{ 1 \}_a) + \dots + (-1)^{b+1}
\frac{\left( \frac{i \pi}{3} \right)^b}{b!}\mu(2,\{ 1 \}_a) \\
&\quad + \int_0^\omega \frac{ (\log (1-z))^b \li_{ \{ 1 \}_{a+1}}(z)}{b! z} dz \\
 &\quad \hspace{.5in}\text{(by repeated integration by parts)} \\
&= \frac{i \pi}{3} \mu(\{ 1 \}_{b-1}, 2, \{ 1 \}_a) + \dots + (-1)^{b+1}
\frac{\left( \frac{i \pi}{3} \right)^b}{b!}\mu(2,\{ 1 \}_a) \\
&\quad+ (-1)^b {a+b + 1 \choose b} \mu(2, \{ 1 \}_{a+b})
\quad \text{using (\ref{eq:lid3})}.
\end{split}
\end{equation*}

Now, multiplying through by $ i^{a+b+2} $, extracting real parts
 and proceeding by induction we find that we must show that:
$$
{a + b + 2 \choose 0} + \dots + (-1)^b { a+b+2 \choose b }
 = (-1)^b {a + b + 1 \choose
b}.
$$
However, $ {a + b + 2 \choose i} = {a + b + 1 \choose i} + {a + b + 1
\choose i-1} $
and it is now easily seen that the left-side telescopes.
\eop

\subsection{Additional Evaluations}

We can use (\ref{eq:gf2}) to obtain a clean expression for the alternating
sum of
all mgl's of the form $\mgl(a, \{ 1 \}_b)$ of fixed weight.
Let
$$ A(x): = \sum_{n=1}^\infty \left( \sum_{m=0}^{n-2} (-1)^{m+1}
\mgl(n+2 -m, \{ 1 \}_m) \right) x^n .
$$
If we  set $ x = y $ in(\ref{eq:gf2}) then, after some work, we obtain
\begin{eqnarray*}
A(x) &=& \frac{1}{2}\left( -\frac{\pi x e^{-\pi x/3}}{\sinh \pi x}
- e^{\frac{\pi}{3}x} + 2 \right) \\
&=&\frac{1}{2}\left( -\frac{2 \pi x e^{2\pi x/3}}{e^{2 \pi x} -1} -
 e^{\frac{\pi}{3}x} + 2 \right).
\end{eqnarray*}
This leads to
\begin{equation}\label{eqn:alt}
 \sum_{m=0}^{n-2} (-1)^{m+1} \mgl(n+2 -m, \{ 1 \}_m)
= \frac{-\pi^n}{2n!} \left( B_n \left( \frac{1}{3} \right) 2^n +
\left( \frac{1}{3} \right)^n \right).
\end{equation}

We note that
(\ref{eqn:alt}) is equivalent to
$$
Re \biggl( 1 - {}_2\mathrm{F}_1(-ix, -ix; 1 - ix; \omega) \biggr)
=\\
 Re \biggl( \sum_{n=1}^\infty \frac{(-ix)(-ix)_n}{n - ix} \omega^n \biggr)
  = - \frac{1}{2} \biggl( \frac{\pi x e^{-\pi x/3}}{\sinh (\pi x)}
   + e^{\frac{\pi x}{3}} - 2 \biggr),
$$
on using the hypergeometric representation of the underlying generating
function.
We have not managed to prove this by more direct methods.

\vspace{\baselineskip}

An unusual looking class of identities may be extracted from (\ref{eq:gfo})
 on setting $ y = 1 - x $.  This gives:
$$
1 - e^{-i\pi x /3} = \sum_{a,b \ge 0} \mu(a + 2, \{1 \}_b) x^{a+1}
(1-x)^{b+1}
$$
which -- when we extract the coefficients of various powers of
 $ x $ on both sides -- gives us curious infinite sums of MCVs of
 different weight
 (reminiscent of similar rational $\zeta$-evaluations~\cite{BBC}).
For example, extracting the coefficient of $ x $ yields:
$$
\sum_{b=0}^\infty \mu(2, \{ 1 \}_b) = \frac{i \pi}{3}.
$$
More generally,  we obtain
$$
\sum_{b=0}^\infty \mu(n+1, \{ 1 \}_b)
 - (b+1)\mu(n, \{ 1 \}_b)
 + \dots + (-1)^{n+1} {b + 1 \choose n - 1} \mu(2, \{ 1 \}_b)
 =-\frac{(-i \pi/3)^n}{n!}.
$$

\section{MCV Dimensional Conjectures}

While there do not appear to be many other closed form
evaluations, it is apparent that there is still more to be learned by
 examining all MCVs -- and especially their integral
 representations. This is a subject
 we have largely ignored in this paper, but which figures large
  in~\cite{sixth}, where polylogarithms of the sixth root of unity
  were studied in the context of integrals arising
  from quantum field theory.

 Experiments using linear relation algorithms suggested that
the only MCVs that evaluate
to      rational multiples of powers of $ \pi $
are those already identified,
namely $ \mgl(3,1) $    and
$ \mgl(\{ 1 \}_b , 2, \{ 1 \}_a) $.
Moreover we found no other non-trivial
reduction of an MCV to a single rational
multiple of powers of other MCVs.
Nevertheless,   our integer relation
searches suggested a very simple enumeration of
the basis size for MCVs of a given weight.

Consider the set
$ {\cal C}(n) :=\{ \mcl(a_1, \dots, a_k) : a_1 + \dots + a_k = n \}$
of      all multiple Clausen values
of fixed weight, $n$.
We wish to determine    the smallest set of real numbers such that each
   element of ${\cal C}(n) $
   can be written as a rational linear combination of elements
    from this set.
    This will consist of mcl's of that weight
     or products of lower weight mcl's, mgl's, MZV's and powers of $ \pi $.
     We denote by $I(n)$ the size of the basis for ${\cal C}(n)$.
     Similarly, we denote by $R(n)$ the basis size for multiple Glaisher
     values
     of weight $n$.

Our first conjecture is quite striking:

\begin{Conj}
The following {\em twisted Fibonacci} recursion obtains:
\begin{align*}
R(n) &= R(n-1) + I(n-2) \\
I(n) &= I(n-1) + R(n-2) \\
R(0) &= R(1) = 1 \mbox{ and }
I(0) = I(1) = 0.
\end{align*}
\end{Conj}

A corollary is that $ W(n) := R(n) +    I(n) $,
the total size of a rational basis for
MCV's of weight $n$, should satisfy
$$ W(n) =        W(n-1) + W(n-2),$$
which delightfully gives the Fibonacci sequence.

Looking at things more finely, we examined the number $P(n,k)$
of {\em irreducibles} of weight $n$ and depth $k$, such that
a rational basis at this weight and depth is formed from
a minimum number of irreducibles,
augmented by products of irreducibles of lesser
weight and depth. Again, we are lead to a rather striking
conjecture:

\begin{Conj}
This weight and depth filtration is generated by
$$
\prod_{n > 1, k > 0} (1-x^ny^k)^{P(n,k)} = 1- \frac{x^2y}{(1-x)}.
$$
\end{Conj}

Both conjectures have been intensively
checked by the PSLQ algorithm~\cite{DHJB1}
for $k\leq n\leq 7$. They provide
compelling evidence     that there
 is a great deal of structure to MCVs.
 It seems unlikely that they will be proven soon, since they
 imply, {\sl inter alia}, the irrationality of $ \zeta(n) $
 for    all odd $ n $.
The interested reader has online access to some of the  code we used,
at

\vspace{\baselineskip}

\centerline{http://www.cecm.sfu.ca/projects/EZFace/Java/
}

\vspace{\baselineskip}

\noindent
and in a forthcoming CECM interface for more general integer relation problems.

\section{
Ap\'ery Sums and
the Golden Ladder}

By way of comparison we present results for the alternating binomial
sums
$$A(k):=\sum_{n>0}\frac{(-1)^{n+1}}{{2n\choose n}n^k}.$$
As we now describe, we found that the cases $k=2, 3, 4, 5, 6$
reduce to classical polylogarithms  of  powers of
$$\rho:=\frac{\sqrt5-1}2,$$   the reciprocal
of the golden section.
The ladder that generates these results
extends up to $\zeta(9)$.
Details of {\em polylogarithmic ladder} techniques are to be found in
\cite{L2}.

 The results for $k<5$ were proven by classical methods
(and also obtained by John Zucker, private communication). For $k\ge5$,
we were content to rely on the empirical methods adopted in~\cite{L2},
determining rational coefficients from
high precision numerical computations.
\vspace{\baselineskip}

\underline{At $k=2$} one easily obtains from       Clausen's
 hypergeometric square, given in ~\cite{AS} or \cite{PA}, the  result
$$A(2)=2L^2$$ where $$L:=\log(\rho).$$ Indeed, we found 6 integer relations
between $A(2)$, $\zeta(2)$, and the dilogarithms
$\{{\rm Li}_2(\rho^p)\mid p\in{\cal C}\}$, where
$${\cal C}:=\{1,2,3,4,6,8,10,12,20,24\}$$
generates the corresponding  cyclotomic relations~\cite{L2}.

In general, it is more convenient to work with the set
$${\cal K}_k:=\{L_k(\rho^p)\mid p\in{\cal C}\}$$
of {\em Kummer-type polylogarithms}, of the form
$$L_k(x):=\frac{1}{(k-1)!}\int_0^x\frac{(-\log|y|)^{k-1}dy}{1-y}
=\sum_{r=0}^{k-1}\frac{(-\log|x|)^r}{r!}{\rm Li}_{k-r}(x)$$
where as before ${\rm Li}_k(x):=\sum_{n>0}x^n/n^k$.

\vspace{\baselineskip}

\underline{At $k=3$} one has Ap\'ery's result $$A(3)=\df25\zeta(3).$$
Moreover there are 5 integer relations between ${\cal K}_3$, $L^3$, and
$\zeta(3)$.

\vspace{\baselineskip}

\underline{At $k=4$} we recently proved a result,
using classical polylogarithmic theory,
 which simplifies to
$$A(4)=4\widetilde{L}_4(\rho)-\df12L^4-7\zeta(4)$$
where $$\widetilde{L}_k(x):=L_k(x)-L_k(-x)=2L_k(x)-2^{1-k}L_k(x^2).$$
In fact, there are 5 integer relations between
${\cal K}_4$, $L^4$, $\zeta(4)$ and $A(4)$.
Another simple example is
$$A(4)=\df{16}{9}\widetilde{L}_4(\rho^3)-2L^4-\df{23}{9}\zeta(4).$$

\vspace{\baselineskip}

\underline{At $k=5$} we found 4
empirical integer relations between ${\cal K}_5$, $L^5$,
$\zeta(5)$ and $A(5)$. The simplest result is
$$A(5)=\df52L_5(\rho^2)+\df13L^5-2\zeta(5).$$
More explicitly,  with $\rho:=(\sqrt5-1)/2$, this produces
\begin{eqnarray}
\sum\limits_{k=1}^\infty {\frac{(-1)^{k+1}}{k^5\binom{2k}k}} &=&2\zeta(5)
-\df43\log(\rho)^5+\df83\log(\rho)^3\zeta(2)+4\log(\rho)^2\zeta(3) \nonumber\\
&+&80\sum\limits_{n>0}\left(\frac{1}{(2n)^5}-
\frac{\log(\rho)}{(2n)^4}\right)\rho^{2n},
\end{eqnarray}
which,  along with previous integer relation exclusion bounds
(see, e.g., \cite{BaB}, \cite{BL}),
 helps explain why no `simple' evaluation for $A(5)$
such as those for $S(2), A(3), S(4)$ has ever been found.
\vspace{\baselineskip}

\underline{At $k=6$} we found the empirical relation
$$11\left\{A(6)-\df25\zeta^2(3)\right\}=
144\widetilde{L}_6(\rho)-\df{64}{9}\widetilde{L}_6(\rho^3)
+\df95L^6-\df{2434}{9}\zeta(6)$$
which is the simplest of 3 integer relations between
${\cal K}_6$, $L^6$, $\zeta(6)$ and the combination
$A(6)-\frac25\zeta^2(3)$.

\vspace{\baselineskip}

\underline{At $k=7$} there are 2 integer relations between
${\cal K}_7$, $L^7$, and $\zeta(7)$. There is no result for $A(7)$
from this set; presumably $A(7)$ occurs in combination with
some other weight-7 irreducible, of which $\zeta^2(3)$ was
a harbinger, at $k=6$.

\vspace{\baselineskip}

\underline{At $k=8$} there is a single integer relation.

\vspace{\baselineskip}

\underline{At $k=9$} the ladder terminates with
a single integer relation, namely
\begin{eqnarray*}
2791022262\,\zeta(9)&=&
 15750\,L_9(\rho^{24})
+74277\,L_9(\rho^{20})
-8750000\,L_9(\rho^{12})\\&&{}
-19014912\,L_9(\rho^{10})
-206671500\,L_9(\rho^{8})\\&&{}
+1295616000\,L_9(\rho^{6})
-3180657375\,L_9(\rho^{4})\\&&{}
+4907952000\,L_9(\rho^{2})
-52537600\log^9(\rho).
\end{eqnarray*}
This still falls short of the ladder for $\zeta(11)$, in~\cite{lad11}.
The current record is set by the ladder for $\zeta(17)$ in~\cite{DHJB2},
which extends the weight-16 analysis of Henri Cohen, Leonard Lewin and Don
Zagier~\cite{CLZ}, in the number field of the Lehmer polynomial of
conjecturally smallest Mahler measure.

\vspace{\baselineskip}
\vfill
\noindent {\bf Acknowledgements}.  This research was in part
performed while Joel Kamnitzer was an NSERC Undergraduate Research Fellow
at CECM in 1999 and while David Broadhurst visited CECM. For this and
other direct NSERC support to Dr Borwein we express our gratitude.

We also wish to thank Simon Plouffe and Philippe Flajolet for communications
which triggered some of our explorations, and David Bailey, David
Bradley, Petr Lison\v ek and John Zucker for many useful exchanges.

\raggedright

\end{document}